\documentclass{elsart}
\newcommand{\ltsima} {$\; \buildrel < \over \sim \;$}
\newcommand{\simlt}  {\lower.5ex\hbox{\ltsima}}            
\newcommand{\gtsima} {$\; \buildrel > \over \sim \;$}
\newcommand{\simgt}  {\lower.5ex\hbox{\gtsima}}            

\usepackage{natbib}

\usepackage{graphicx}

\usepackage{amssymb}

\begin{document}

\begin{frontmatter}



\title{The Ultraluminous X-ray Sources \\ NGC 1313 X-1 and X-2}


\author[label1]{Turolla, R.,}
\author[label2,label3]{Mucciarelli, P.,}
\author[label2]{Zampieri, L.,}
\author[label2]{Falomo, R.,}
\author[label4]{Chieregato M.}
\author[label4]{and Treves A.}
\address[label1]{Department of Physics, University of Padova, via Marzolo
8, I-35131 Padova, Italy}
\address[label2]{INAF-Osservatorio Astronomico di Padova, vicolo
dell'Osservatorio 5, I-35122 Padova, Italy}
\address[label3]{Department of Astronomy, University of Padova, 
vicolo dell'Osservatorio 5, I-35122 Padova, Italy}
\address[label4]{Department of Physics and Mathematics, University of 
Insubria at Como, Via Valleggio 11, I-22100 Como, Italy}

\begin{abstract}
We present a detailed analysis of {\it XMM} archival data of two 
Ultraluminous X-ray Sources (ULXs) in the nearby spiral galaxy 
NGC 1313: NGC 1313 X-1 and X-2. The spectral continuum of
these sources was modeled with a soft thermal 
component plus a power-law. If the soft component originates 
from an accretion disk, the inferred mass of the compact remnant 
is $\simgt 100 M_\odot$, making it an Intermediate Mass Black Hole (IMBH). 
A detailed analysis of the residuals of the {\it XMM} EPIC-pn spectrum 
shows some evidence for the presence 
of an Oxygen emission line in NGC 1313 X-1. The simultaneous presence 
of an excess in emission, although at a much reduced significance 
level, at different energies in the X-ray spectra of NGC 1313 X-1 and 
X-2 is suggestive of typical emission lines from young supernova 
remnants. An optical counterpart for NGC 1313 X-2 was also identified. On 
an ESO 3.6 m image, the {\it Chandra} error box embraces a $R\sim 23$ mag
stellar-like object and excludes a previously proposed optical
counterpart. 
\end{abstract}

\begin{keyword}
Black holes \sep 
X-rays: individual (NGC 1313 X-1, NGC 1313 X-2/MS 0317.7-6647) \sep 
X-rays: binaries \sep X-rays: galaxies
\end{keyword}

\end{frontmatter}

\section{Introduction}
\label{}
Point-like, off-nuclear X-ray sources with luminosities in excess of
the Eddington limit for one solar mass are increasingly discovered in
nearby galaxies. Despite the growing body of observational data, the
mechanism which powers ULXs is still under debate. In several cases 
a variability on timescales of months/years has been detected, hinting 
towards the presence of a compact object. 
Were these sources X-ray binaries in the host galaxy, assuming 
Eddington-limited accretion, masses in the range
$M_{\mathrm{BH}}\sim 100-1000 M_\odot$ are inferred for the compact object from
the observed flux. IMBH may have
originated from the collapse of a massive star formed in a low
metallicity environment, or through merging of massive stars (or lower
mass BHs) in a cluster \citep{mc03}. The BH
mass estimates derived from considerations involving the Eddington
limit are, however, questionable because these sources need not to be
spherically symmetric, nor stationary. It has been proposed that many
of the ULX properties can be explained assuming that they do not emit
isotropically \citep{king01} or are dominated by emission
from a relativistic jet \citep{kaaret03}. In this case, they 
may harbor stellar mass BHs and may be similar to Galactic 
microquasars.
\\
Particularly interesting  is the case of the three ULXs
hosted in the nearby SBc galaxy NGC 1313. One of them is known 
to be associated with the interacting supernova SN 1978K \citep{sch00}. 
The other two have been extensively
studied with several X-ray telescopes and, to date,
provide some of the best evidence for the presence of a soft component
in the X-ray spectrum of ULXs. Moreover, as discussed in \cite{mil04}
 and \cite{zamp04}, hereafter Z04, their continuum 
subtracted {\it XMM} EPIC-pn spectra show significant residuals 
especially in the 0.5--3 keV range. 

\section{X-ray Data}
{\it XMM-Newton} observed NGC 1313 on October 17, 2000 for a total of
$\sim$ 42 Ks. The field was centered on the galaxy nucleus and
contains all the three ULXs. The three {\it XMM} EPIC cameras 
operated in Prime Full Window mode with the medium filter. The
analysis reported here follows that presented in Z04. 
We consider only the EPIC-pn spectrum 
because it has twice as many counts as each single MOS instrument. 
The EPIC-pn spectrum was directly extracted from the observation
data file because the automatic pipeline processing failed to produce 
an event list. Data screening, region selection and
event extraction were performed with the standard software {\sc
XMM-SAS} v 6.0.0. In order to
eliminate the possible contamination of solar flares (present during
the observation), event files were filtered using the good time
intervals when the total off-source count rate above 10 keV is less
than 1 counts s$^{-1}$. This leave $\sim 21$ ks  of useful time 
with an average count rate of $0.73$ counts s$^{-1}$ for NGC 1313 X-1
and $0.24$ counts s$^{-1}$ for NGC 1313 X-2.

The best fit of the {\it XMM} EPIC-pn continuum for both sources is 
obtained 
with an absorbed soft, thermal component (a multicolor disk blackbody, MCD) 
plus a power-law. In comparison with previous {\it ROSAT} and {\it ASCA\/} 
data the 
statistics is sufficiently good that two components models provide a 
significant improvement over single component ones (see Z04). The statistical 
improvement obtained adding a MCD component to a power-law model is 
significant above the $4.5 \sigma$ level.  The resulting 
best fitting parameters for NGC 1313 X-1 and X-2 are 
reported in Table \ref{tab1}. The inferred inner disk temperature is much lower
than that obtained for single component thermal models, as already
found by \cite{mil02}. The X-ray luminosity in the 0.2--10 keV range,
assuming isotropic emission and a distance of the host galaxy 
$\simeq 3.7$ Mpc, is $\simeq (1.4 \pm 0.2) \times 10^{40}$erg s$^{-1}$ for 
NGC 1313 X-1 and $\simeq (6.0 \pm 0.5) \times 10^{39}$erg s$^{-1}$ for 
NGC 1313 X-2. If at maximum the source radiates at the 
Eddington limit, the BH mass is $\sim 120 \, M_\odot$ for X-1 and 
$\sim 50 \, M_\odot$ for X-2. Sub-Eddington accretion would 
imply an even larger mass.

Residuals in the EPIC-pn spectra of NGC 1313 X-1 and X-2 suggest the presence 
of some emission lines in both sources. 
In order to test for the presence of spectral features,
we added, one by one, gaussian components at different energies where
residuals show evidence of some excess in emission.

For NGC 1313 X-1, the most significant residual is at around 0.6 keV,
identified with a high ionization (He-like) Oxygen line (see Figure 
1). Adding a gaussian component, the improvement of the fit
with respect to an absorbed MCD+PL model is significant at the 3.5
$\sigma$ level. The line energy is 0.58 keV, while the line
width was frozen in the fit (0.01 keV). Residuals show also an excess
of emission at energies of 1.8 and 4.7 keV (see Figure 1). 
The first is the typical energy of highly ionized
Silicon, while the second is not readily identified. Adding other two
gaussian components for fitting these residuals does not give any
further statistical improvement in the fit. 

A similar analysis was performed also on the EPIC-pn spectrum of NGC
1313 X-2, following the same approach outlined above. Also in this
case the spectrum shows some residuals at an energy of 0.6 keV. 
In order to test the influence of ISM absorption models (due to the 
neutral Oxygen absorption edge, see \citealt{mil04}) on the detection of 
an O line in NGC 1313 X-1, we repeated our analysis using the T\"ubingen-Boulder
absorption model (TBABS; \citealt{wil00}, Table \ref{tab1}), that includes 
a
treatment of gas-phase ISM with revised photoionization cross sections
and revised abundances (plus the contribution of grain-phase ISM and
molecules). The improvement obtained adding a gaussian component to 
a TBABS+MCD+PL continuum is at the $\sim 3.6 \, \sigma$ level. 
The line energy is 0.6 keV, while the line width was
frozen in the fit (0.01 keV). The significance of the line remains
fairly high even performing a fit in a restricted energy region
(0.3--1.0 keV) around the centroid of the gaussian. Fitting the
continuum with an absorbed MCD+PL model, the addition of a gaussian
component with fixed width (0.01 keV) gives an improvement at the
$\sim 3 \, \sigma$ level.

Additional information on the line properties were obtained analyzing
the line profiles. After freezing the best fitting continuum of Table 
\ref{tab1} and removing the gaussian components,
residuals (normalized to the continuum) were fitted with a constant
(equal to unity) plus a gaussian profile, obtaining the lines centroid
energies and equivalent widths. The energies are consistent (within
the errors) with those derived from the spectral fits, while the
equivalent widths are all below 100 eV. A thorough analysis of these 
data and new XMM observations of NGC 1313 X-1 and
X-2 will be presented in Mucciarelli et al. (2005, in preparation).

\section{The optical counterpart of NGC 1313 X-2}
Optical images of the field of NGC 1313 X-2 in the $R$-band were taken 
on 16 January 2002 with the 3.6 m telescope of the European Southern 
Observatory (ESO) at La Silla (Chile). Four images were obtained for a 
total exposure time of 1320 s (see Z04 for details). The accuracy  of the
astrometric calibration, performed with GSC2 ESO field stars, is 0.3$"$ 
(1-$\sigma$). The X-ray position of NCG 1313 
X-2 was obtained from a 2002 {\it Chandra} pointing with an accuracy 
of 0.7$"$ (Z04). The astrometric calibration of the optical and {\it Chandra}
images was checked  using the very accurate radio position 
of SN 1978K (\citealt{ryd93}, Z04).
Our {\it Chandra} position of NGC 1313 X-2 is shown in
Figure \ref{fig2}, together with the {\it ROSAT\/} HRI \citep{sch00}
and {\it XMM} EPIC \citep{mil02} error boxes, overlaid on our ESO
image. All measurements are consistent within 1-$\sigma$. Object 
C ($R\sim 23$)
is inside the Chandra error box and its position coincides within
1-$\sigma$ with that of NGC 1313 X-2, making it a likely counterpart 
and ruling out the previous proposed counterpart, object A.
From the maximum absorbed X-ray flux of NGC 1313 X-2 ($f_X\sim 2\times
10^{-12}$ erg cm$^{-2}$ s$^{-1}$) and optical magnitude of object C, 
we estimate $f_X/f_{opt}\simgt 1000$. 

\section{Discussion}
The X-ray spectral parameters, in particular the temperature of the MCD fit 
($T_{MCD}$), can be used to estimate the BH mass. 
From the effective temperature of a standard accretion disk, under the
assumption  that $T_{MCD}$ represents an estimate of the maximum 
disk temperature, Z04 infer  $M_{BH}/M_\odot=({\dot M}c^2/L_{Edd}) f^4
(\alpha T_{MCD}/1.5\times 10^7 \, {\rm K})^{-4}$ (where $f\sim 1.5$ is 
a color correction
factor). The temperature obtained from the two-components fit to
the {\it XMM} spectrum (Table \ref{tab1}) implies $M_{BH}
\approx (130/90) f^4 \, M_\odot$ for NGC 1313 X-1/X-2
(assuming $\alpha\simeq 2$ and Eddington limited
accretion).

Up to now evidence of emission lines in ULXs has been reported in M82 X-1
\citep{stro03}, NGC 4559 X-10 \citep{crop04} and M101 ULX-1 
\citep{kong05}. 
The presence of an
Oxygen line in the EPIC-pn spectrum of NGC 1313 X-1 appears to be
significant at above $3.5 \, \sigma$ while the statistical evidence of
other features is at most marginal ($< 2 \, \sigma$). 
The simplest interpretation is that we are observing 
typical emission features of intermediate mass elements left over 
after the explosion 
of one or more supernovae belonging to the stellar association 
of massive stars where the two ULXs are probably embedded.

Concerning NGC 1313 X-2, the luminosity inferred from the 
apparent magnitude of object C is 
consistent with a $\approx 20 M_\odot$ 
main sequence star or a $\sim 15-20 M_\odot$ evolved OB supergiant
making NGC 1313 X-2 a High-Mass X-ray Binary. 
This picture
seems to be confirmed also by the study of the environment of NGC 1313 X-2 (see Z04)
and by the association of both ULXs with extended optical emission nebulae 
\citep{pak02}.


\bigskip

\begin{table*}[ht!]
\scriptsize
\centering
\caption{Parameters of the fit of the {\it XMM}-EPIC observation
of NGC 1313 X-1 and X-2}
\label{tab1}
\vspace{0.1in}
\begin{tabular}{l l l l l l }
\hline
\hline
Model & $\frac{N_{\mathrm{H}}}{10^{21} cm^{-2}}$ & $\Gamma$ & $kT$ [keV] &
Parameters & $\chi^2_{\mathrm{red}}$(dof)\\
\hline
\multicolumn{6}{c}{NGC 1313 X-1}\\
\hline
WABS(MCD+PL)		& $3.48_{-0.25}^{+0.60}$	&
$1.75_{-0.04}^{+0.08}$          & $0.19_{-0.03}^{+0.01}$     & & 1.12(301) \\
TBABS(MCD+PL)		& 4.21$_{-0.32}^{+0.64}$	& 
1.69$_{-0.05}^{+0.07}$  &0.19$_{-0.02}^{+0.01}$ &&1.10(301)\\
TBABS(MCD+PL+GAUSS)	& 4.48$_{-0.31}^{+0.73}$	& 
1.79$_{-0.05}^{+0.06}$	&0.17$_{-0.02}^{+0.01}$	&$E=0.60_{-0.02}^{+0.02}$
keV &1.04(299)\\
\hline
\hline
\multicolumn{6}{c}{NGC 1313 X-2}\\
\hline
WABS(MCD+PL)                    & $3.26_{-0.25}^{+0.92}$        &
$2.17_{-0.09}^{+0.15}$          & $0.20_{-0.04}^{+0.05}$     & & 1.21(109)\\
\hline
\end{tabular}
\end{table*}
\vspace{0.2in}

\begin{figure}[hb!]
\label{figx1}
\begin{center}
\includegraphics[width=3.1in,angle=-90]{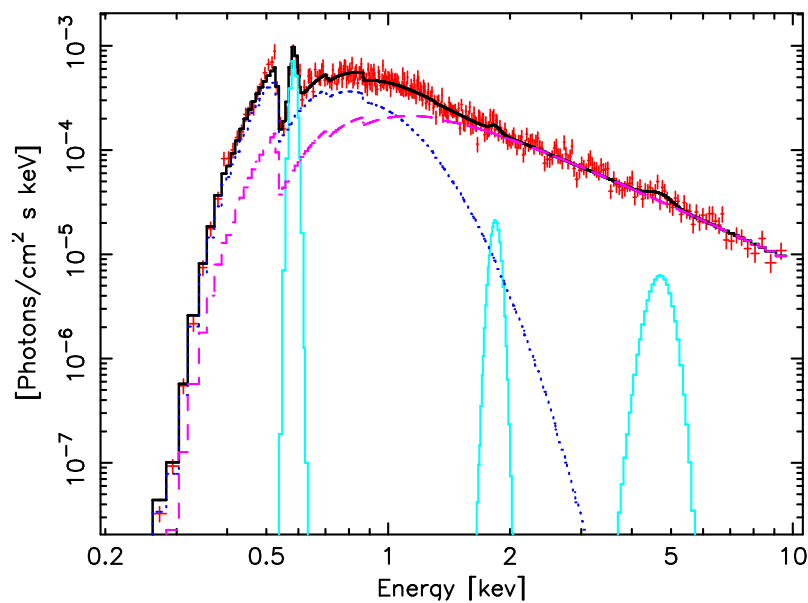}
\end{center}
\caption{{\it XMM} EPIC-pn spectrum of NGC 1313 X-1:
absorbed MCD (dotted line)+PL (dashed line) model with three gaussian
components (thin solid lines) at energies 0.59, 1.8 and 4.7 keV. 
}
\end{figure}
\vspace{0.2in}

\begin{figure}[hb!]
\begin{center}
\includegraphics[height=3.1in]{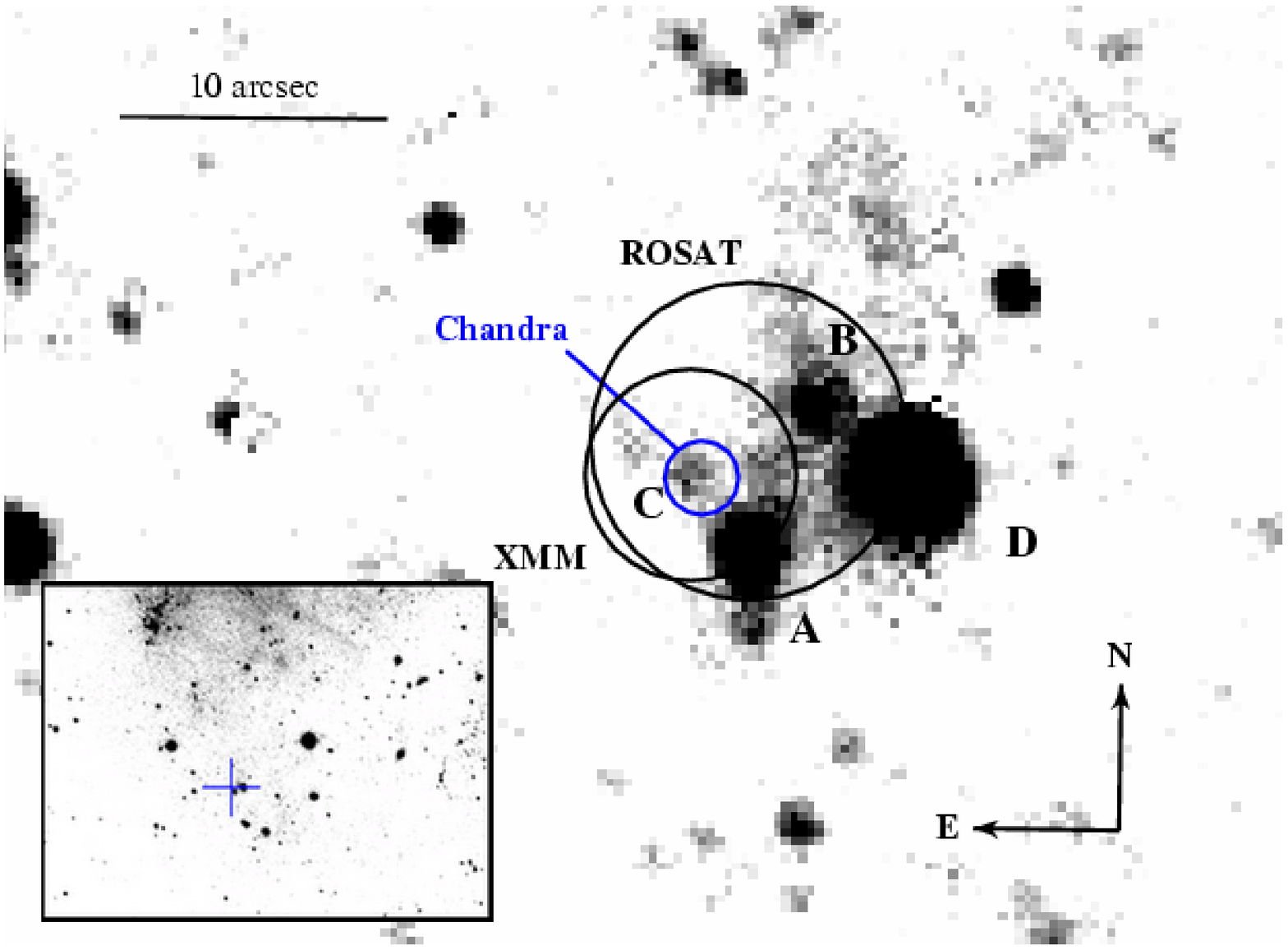}
\end{center}
\caption{ESO 3.6m $R$-band image of the field of NGC 1313 X-2.
The circles show the {\it ROSAT\/} HRI, {\it XMM\/} EPIC and {\it
Chandra} ACIS-S positions. The estimated 90\% confidence radii are
6$''$, 4$''$ and 1.4$''$ respectively. Labels A and
C mark the old and new proposed optical counterparts. The insert at 
the bottom-left shows a larger portion of the
image with the position of the X-ray source (cross).}
\label{fig2}
\end{figure}


\footnotesize

\begin{thebibliography}{}

\bibitem[\protect\citeauthoryear{Cropper et al.}{2004}]{crop04}
Cropper, M. et al. 2004,  MNRAS, 349, 39-51, {\em Probable 
intermediate-mass 
black holes in NGC 4559: XMM-Newton spectral and timing constraints}
\bibitem[\protect\citeauthoryear{Kaaret et al.}{2003}]{kaaret03}  
Kaaret, P., Corbel, S., Prestwich, A.H., \& Zezas, A. 2003, Science,
299, 365-368, {\em Radio Emission from an Ultraluminous X-ray Source}
\bibitem[\protect\citeauthoryear{King et al.}{2001}]{king01}
King, A.R., et al. 2001, ApJL, 552, L109-L112, {\em Ultraluminous X-Ray 
Sources in External Galaxies}
\bibitem[\protect\citeauthoryear{Kong et al.}{2005}]{kong05} 
Kong, A.K.H., Rupen, M.P., Sjouwerman, L.O., \& Di Stefano, R.
2005, paper presented at the XXII Texas Symposium on Relativistic Astrophysics, 
Stanford University, Dec. 13-17, 2004  (astro-ph/0503465), {\em 
Multiwavelength Observations and State Transitions of an Ultra-luminous 
Supersoft X-ray Source: Evidence for an Intermediate-Mass Black Hole} 
\bibitem[\protect\citeauthoryear{Miller \& Colbert}{2004}]{mc03}
Miller, M.C., \& Colbert, E.J.M., 2004, Int. Jour. of Modern
Physics, 13, 1-64, {\em Intermediate-Mass Black Holes}
\bibitem[\protect\citeauthoryear{Miller et al.}{2004}]{mil04}
Miller, J.M., Fabian, A.C., \& Miller, M.C. 2004, ApJ, 607, 931-938,
{\em Revealing a Cool Accretion Disk in the Ultraluminous X-Ray Source 
M81 X-9 (Holmberg IX X-1): Evidence for an Intermediate-Mass Black Hole}
\bibitem[\protect\citeauthoryear{Miller et al.}{2003}]{mil02} 
Miller, J.M., Fabbiano, G., Miller, M.C., \& Fabian, A.C. 2003, ApJL, 585, 
L37-L40, {\em X-Ray Spectroscopic Evidence for Intermediate-Mass Black 
Holes: Cool Accretion Disks in Two Ultraluminous X-Ray Sources}
\bibitem[\protect\citeauthoryear{Pakull \& Mirioni}{2002}]{pak02} 
Pakull, M.W., \& Mirioni, L. 2002, in Proc. ESA Symp., New Visions of the X-ray
Universe in the {\it XMM-Newton\/} and {\it Chandra\/} Era, eds.
F. Jansen et al. (ESA SP-488) (astro-ph/0202488), {\em Optical 
Counterparts of Ultraluminous X-ray Sources}
\bibitem[\protect\citeauthoryear{Ryder et al.}{1993}]{ryd93}
Ryder, S. et al. 1993, ApJ, 416, 167-181, {\em SN 1978K: an Extraordinary 
Supernova in the Nearby Galaxy NGC 1313}
\bibitem[\protect\citeauthoryear{Schlegel et al.}{2000}]{sch00} 
Schlegel, E.M., Petre, R., Colbert, E.J.M., \& Miller, S. 
2000, AJ, 120, 2373-2382, {\em A Deep ROSAT HRI Observation of NGC 1313}
\bibitem[\protect\citeauthoryear{Strohmayer et al.}{2003}]{stro03} 
Strohmayer, T.E., \& Mushotzky, R.F., 2003, ApJL, 586, L61-L64, 
{\em Discovery 
of X-Ray Quasi-periodic Oscillations from an Ultraluminous X-Ray Source 
in M82: Evidence against Beaming} 
\bibitem[\protect\citeauthoryear{Wilms et al.}{2000}]{wil00} 
Wilms, J., Allen, A., \& McCray, R., 2000, ApJ, 542, 914-924, {\em On the 
Absorption of X-Rays in the Interstellar Medium }
\bibitem[\protect\citeauthoryear{Zampieri et al.}{2004}]{zamp04} 
Zampieri, L., et al. 2004, ApJ, 603, 523-530, {\em The Ultraluminous 
X-Ray 
Source NGC 1313 X-2 (MS 0317.7-6647) and Its Environment}
\end{thebibliography}
\end{document}